\documentclass[twocolumn,english,aps,prb,twocolumn]{revtex4}
\usepackage[T1]{fontenc}
\usepackage[latin9]{inputenc}
\usepackage{amsmath}
\usepackage{amssymb}
\usepackage{graphicx}
\usepackage{esint}

\makeatletter
\@ifundefined{textcolor}{}
{%
 \definecolor{BLACK}{gray}{0}
 \definecolor{WHITE}{gray}{1}
 \definecolor{RED}{rgb}{1,0,0}
 \definecolor{GREEN}{rgb}{0,1,0}
 \definecolor{BLUE}{rgb}{0,0,1}
 \definecolor{CYAN}{cmyk}{1,0,0,0}
 \definecolor{MAGENTA}{cmyk}{0,1,0,0}
 \definecolor{YELLOW}{cmyk}{0,0,1,0}
 }

\usepackage{babel}

\usepackage{babel}

\usepackage{babel}

\usepackage{babel}

\makeatother

\usepackage{babel}
\begin{document}

\title{Valley-kink in Bilayer Graphene at $\nu=0$: A Charge Density Signature
for Quantum Hall Ferromagnetism}

\author{Chia-Wei Huang}

\affiliation{Department of Physics, Bar-Ilan University, Ramat Gan, 52900, Israel}

\author{Efrat Shimshoni}

\affiliation{Department of Physics, Bar-Ilan University, Ramat Gan, 52900, Israel}

\author{H. A. Fertig}

\affiliation{Department of Physics, Indiana University, Bloomington, IN 47405,
USA}

\date{\today}
\begin{abstract}
We investigate interaction-induced valley domain walls in bilayer
graphene in the $\nu=0$ quantum Hall state, subject to a perpendicular
electric field that is antisymmetric across a line in the sample.
Such a state can be realized in a double-gated suspended sample, where
the electric field changes sign across a line in the middle. The non-interacting
energy spectrum of the ground state is characterized by a sharp domain
wall between two valley-polarized regions. Using the Hartree-Fock
approximation, we find that the Coulomb interaction opens a gap between
the two lowest-lying states near the Fermi level, yielding a smooth
domain wall with a kink configuration in the valley index. Our results
suggest the possibility to visualize the domain wall via measuring
the charge density difference between the two graphene layers, which
we find exhibits a characteristic pattern. The width of the kink and
the resulting pattern can be tuned by the interplay between the magnetic
field and gate electric fields. 
\end{abstract}
\maketitle

\section{Introduction}

Two dimensional electron systems in magnetic fields exhibit a great
richness of physics, particularly in the high field regime where the
decreasing radius of the cyclotron orbits gives rise to increasing
importance of electron-electron interactions. Two examples are the
fractional quantum Hall effect (FQHE) and quantum Hall ferromagnets
in the integer QHE.\cite{Sarma1997,Prange1990} The essential feature
of the former is a condensation of the electrons into unusual correlated
states which minimize the Coulomb energy, allowing the electrons to
avoid each other as much as possible. Similarly, for the latter, Coulomb
interactions induce nonperturbative effects on the highly degenerate
Landau bands of the non-interacting system. In particular, due to
exchange, ferromagnetism is induced in the system. A prominent manifestation
of this state is the formation of Skyrmions as novel low energy excitations
of the spin-polarized ground state (or isospin polarized states in
bilayer QH systems).\cite{Prange1990,Sarma1997,Sondhi1993,Fertig1994,Moon1995}

Quantum Hall ferromagnets have also been predicted for graphene in
the integer quantum Hall regimes,\cite{Abanin2007,Bolotin2009,Du2009,Fertig2006,Zhao2010,Barlas2008,Cote2008}
which exhibit particle-hole conjugate Landau levels and a peculiar
$\nu=0$ QH state at zero energy.\cite{Novoselov2005,Novoselov2006,Zhang2005,Zheng2002}
These two unique features are manifestations of the Dirac equation
which governs the electron dynamics near the $\mathbf{K}$ and $\mathbf{K'}$
points in the band structure. For non-interacting electrons in graphene,
four Landau levels are present near zero energy, associated with the
two valleys and the two spin states. In this situation the Zeeman
coupling separates the states into two pairs above and below the Fermi
energy. When interactions are included, the half-filled zero energy
states spontaneously polarize due to exchange and give rise to a ferromagnetic
ground state,\cite{Sarma1997} which may be spin or valley polarized
depending on the strength of the field.\cite{Zhang2006,Zhao2012}

In addition to this interesting bulk property in the $\nu=0$ state,
a coherent domain wall\cite{Fertig2006,Shimshoni2009} (DW) will be
present between a spin polarized bulk state and an unpolarized region
at the physical edge of a finite graphene ribbon.\cite{CastroNeto2006,Brey2006}
This DW has also been predicted to support a Luttinger liquid edge
mode, which is another manifestation of the Coulomb interaction in
2D systems. However, it may be difficult to realize this spin configuration
in currently available graphene ribbons, as their edges are in general
rough.\cite{Li2011} Moreover, such a pattern in the ground state
is hard to probe directly.

\begin{figure}
\includegraphics[width=1\columnwidth]{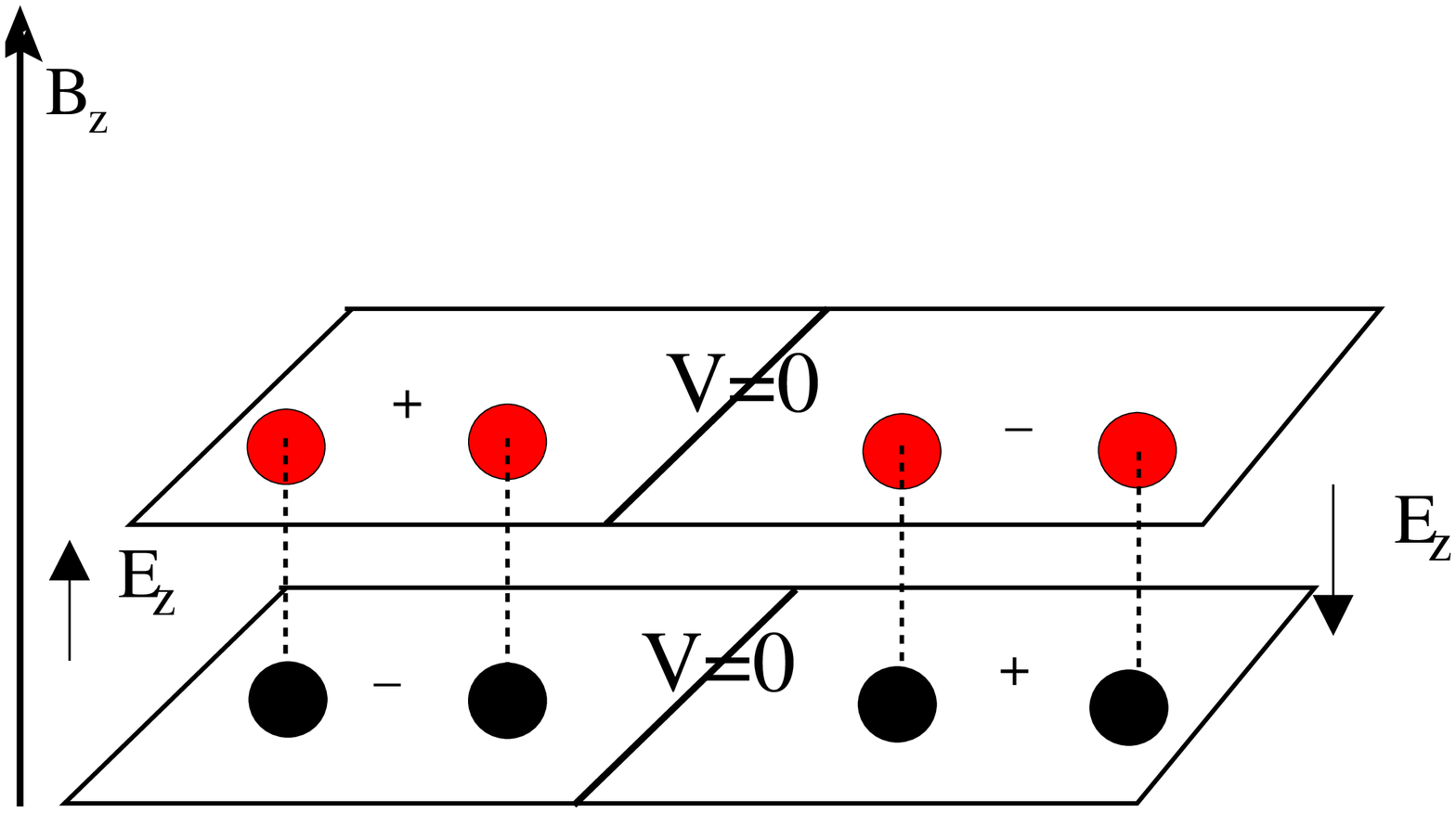}\vspace{8mm}
 \includegraphics[width=1\columnwidth]{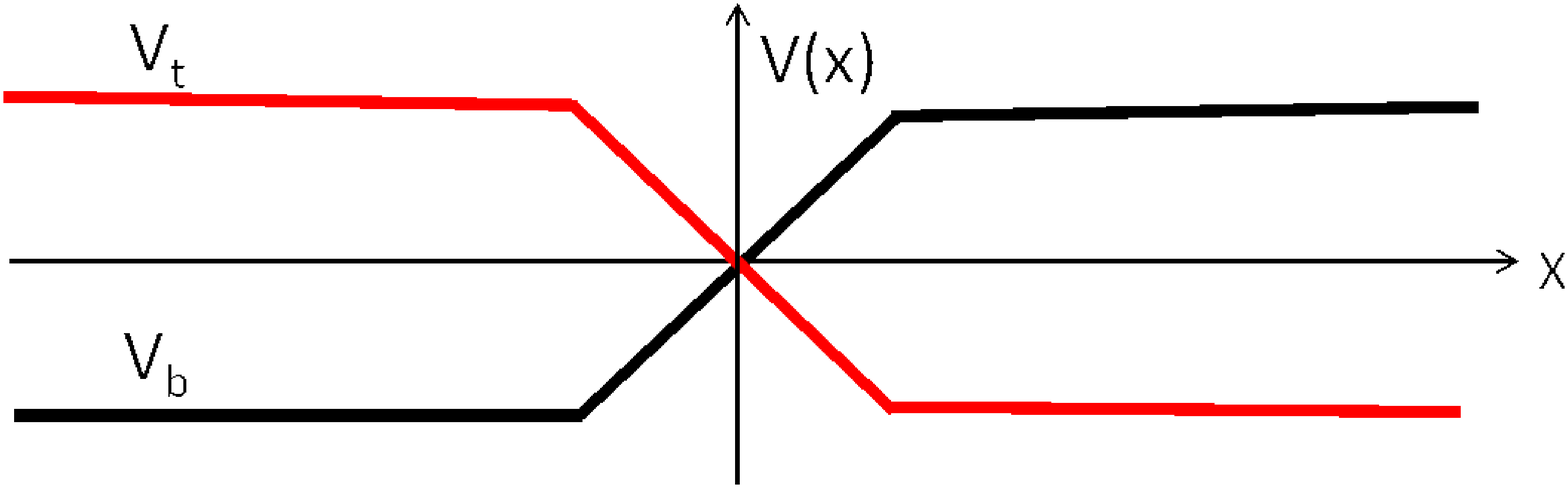} \caption{\label{fig:topo}(Color online) Bilayer graphene in a double gated
system. Upper panel: the red dot ($B$) represents the sublattice
$B$ on the top layer, and the black dot ($\tilde{A}$) represents
the sublattice $A$ on the bottom layer of a BLG. The imaginary line
connecting the $\tilde{A}B$ dimmer represents $\gamma_{1}$ bonding.
The polarities are different on the two sides of the system. Lower
panel: $V_{t}$ marked by red solid line represents the potential
profile for top layer, and $V_{b}$ marked by black solid line represents
the potential profile for the bottom layer. Here we assume an adiabatic
linear profile across the middle of the sample.}
\end{figure}

An {}``internal edge'' in biased bilayer graphene (BLG) proposed
by Martin $\mathit{et\:}al.$\textit{ }provides an alternative way
to create a DW that circumvents the difficulty in making perfect physical
edges.\cite{Martin2008,Zarenia2011} This clean edge can be created
in the middle of a bilayer graphene sample by placing it in an electric
quadrupole gate where a potential profile changes sign across the
center of the sample, as shown in Fig. \ref{fig:topo}. When the Fermi
level is placed at zero energy, a pair of surface states with opposite
chiralities and opposite isospins (valley index) are formed in the
middle of the sample. These states are localized and resemble the
edge states of quantum Hall systems near a physical edge.

In the QH regime, bilayer graphene also exhibits particle-hole symmetric
LLs and particle-hole degenerate zero energy states. Relative to the
monolayer, the layer degrees of freedom of the bilayer system doubles
the zero energy degeneracy. Perpendicular electric fields act as Zeeman
fields for the layer degrees of freedom,\cite{McCann2006,McCann2006a}
thus lifting their degeneracy. This effective Zeeman field can be
tuned to be much larger than the Zeeman splitting of the real spin,
set by the magnetic field. In the double-gated setting, this isospin
Zeeman splitting changes sign in the middle of the sample, yielding
level crossings similar to the physical edge of a monolayer graphene
sample. When interactions are included, QH ferromagnetism sets in
and the fully polarized ground state acquires a finite spin stiffness.
As a result, a coherent DW analogous to the spin DW found in Ref.
\onlinecite{Fertig2006} may form.

In this paper, we study the interaction-induced valley DW in bilayer
graphene in the $\nu=0$ quantum Hall state, in a physical configuration
as shown in Fig. \ref{fig:topo}. The perpendicular magnetic field
$B_{z}$, the strength of the perpendicular bias $E_{z}$, and the
separation of two electric gates are controllable parameters. We use
the Hartree-Fock approximation to derive the ground state and to evaluate
the width of the DW in terms of these parameters. We find that the
DW has an interlayer charge density difference pattern, which may
be accessible experimentally.

This paper is organized as follows. In Section \ref{sec:Continuum-description},
we review the non-interacting energy spectrum of bilayer graphene
with Bernal stacking under a perpendicular magnetic field and a double
gated bias with different polarities. In Section \ref{sec:Coherence-domain-walls},
we derive the ground state wavefunction and energy of the valley-kink
domain wall within a self-consistent Hartree-Fock approximation, and
evaluate the size of the coherent domain wall. Section \ref{sec:Interlayer-charge-density}
discusses the resulting charge density pattern. Finally, we summarize
our results and discuss future directions in Section \ref{sec:Concluding-remarks}.

\section{\label{sec:Continuum-description}Non-interacting model}

We consider a bilayer graphene sheet subject to a perpendicular electric
field which varies along the $x$ direction as shown in Fig. \ref{fig:topo}.
With a gauge choice of $\mathbf{A}=\widehat{y}xB_{z}$ for the magnetic
vector potential, the electron wavefunctions are localized in the
$x$ direction and extended along the $y$ direction with a good quantum
number $k_{y}$. When the Zeeman splitting is small, the low-energy
Hamiltonian of biased bilayer graphene with Bernal stacking in the
vicinity of the $\mathbf{K}$ valley is\cite{Martin2008,Mazo2011}
\begin{equation}
H_{\mathbf{K}}=\left(\begin{array}{cccc}
-V(x)/2 & \omega_{c}a & 0 & 0\\
\omega_{c}a^{\dagger} & -V(x)/2 & \gamma_{1} & 0\\
0 & \gamma_{1} & V(x)/2 & \omega_{c}a\\
0 & 0 & \omega_{c}a^{\dagger} & V(x)/2
\end{array}\right),\label{eq:BGL}
\end{equation}
 where the basis for the Hamiltonian is $\left(\tilde{\left\langle B\right|},\tilde{\left\langle A\right|},\left\langle B\right|,\left\langle A\right|\right)^{\dag}$,
and $A$, $B$ ($\tilde{A}$, $\tilde{B}$) represent the sublattice
wavefunctions on the top and bottom layer respectively. Here $a=\left[\partial_{x}+(x-X)\right]/\sqrt{2}$
and $a^{\dagger}=\left[-\partial_{x}+(x-X)\right]/\sqrt{2}$, where
$x$ (and all length scales henceforth) is in units of the magnetic
length $l_{B}$, the guiding center is defined as $X=l_{B}k_{y}$,
and $\omega_{c}=\sqrt{2}\hbar v_{F}/l_{B}$ $(v_{F}\thickapprox10^{6}ms^{-1})$.\cite{CastroNeto2009,Wallace1947}
The dominant interlayer coupling constant ($\gamma_{1}\sim0.3$ eV)
included in the model is between the $\tilde{A}B$ dimer sites. $V(x)$
is the interlayer bias, assumed to be adiabatically varied so that
in the effective Hamiltonian for a given $x$, $V(x)$ may be replaced
by $V(X)$. For simplicity, we consider 
\begin{eqnarray}
V(X) & = & \begin{cases}
-V, & X<-w/2\\
\left(\frac{V}{w}\right)X, & -w/2<X<w/2\\
V, & X>w/2
\end{cases},\label{eq:Vprofile}
\end{eqnarray}
 where $w$ defines the separation of the two electric gates with
opposite polarities, assumed to be much larger than $l_{B}$. The
two lowest lying eigenvalues are 
\begin{equation}
\varepsilon_{1}^{\mathbf{K}}=\frac{V(X)}{2}\label{eq:e1}
\end{equation}
 and 
\begin{equation}
\varepsilon_{2}^{\mathbf{K}}=\frac{\gamma_{1}^{2}-\omega_{c}^{2}}{2\left(\gamma_{1}^{2}+\omega_{c}^{2}\right)}V(X),\label{eq:e2}
\end{equation}
 and their corresponding eigenstates are 
\begin{equation}
\phi_{1,\mathbf{K}X}(\mathbf{r})=\frac{e^{i\left(K+X\right)y}}{\sqrt{L_{y}}}\left(\begin{array}{c}
0\\
0\\
0\\
\Phi_{0}(x-X)
\end{array}\right),\label{eq:phi1}
\end{equation}

\begin{equation}
\phi_{2,\mathbf{K}X}(\mathbf{r})=\frac{e^{i\left(K+X\right)y}}{\sqrt{L_{y}}N_{h}}\left(\begin{array}{c}
0\\
\Phi_{0}(x-X)\\
\frac{\gamma_{1}V(X)}{\gamma_{1}^{2}+\omega_{c}^{2}}\Phi_{0}(x-X)\\
-\frac{\gamma_{1}}{\omega_{c}}\Phi_{1}(x-X)
\end{array}\right),\label{eq:phi2}
\end{equation}
 where $\Phi_{n}(x-X)$ are the harmonic oscillator wavefunctions
and $N_{h}$ is a normalization factor. The $\phi_{1}$ state is purely
from the lowest Landau level (LLL) of the top layer, while the $\phi_{2}$
state consists of the LLL from the bottom layer and the first LL from
the top layer. Their distinction will be clear when we calculate the
exchange energies. The representation of the Hamiltonian in $\mathbb{\mathbf{K'}}$
has the same form with $V(X)\rightarrow-V(X)$, and a basis where
the order of components in the 4-spinor is inverted, hence $\varepsilon_{i}^{\mathbf{K'}}(X)=-\varepsilon_{i}^{\mathbf{K}}(X)$
(with $i=1,2$).\cite{Mazo2011} Note this means that the low-lying
states of the $\mathbf{K}$ valley reside primarily in one layer,
while those of the $\mathbf{K^{\prime}}$ valley are primarily in
the other. 
\begin{figure}
\begin{centering}
\includegraphics[width=1\columnwidth]{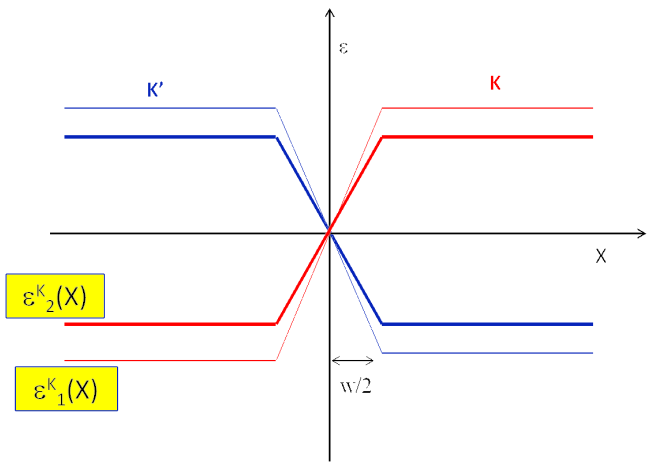} 
\par\end{centering}

\caption{\label{fig:Quantum-Hall-energy}Quantum Hall energy spectrum of a
double gated BLG, as a function of guiding center. We assume a smooth
change in bias across $x=0$, and it yields the existence of zero
energy states at $V=0$.}
\end{figure}

As shown in Eqs. (\ref{eq:e1}) and (\ref{eq:e2}), the energy eigenvalues
$\varepsilon_{i}^{\mathbf{K}}(X)$ and $\varepsilon_{i}^{\mathbf{K'}}(X)$
are determined by the profile of $V(X)$. They all vanish at $X=0$
where there is a level crossing between the $\mathbf{K}$ and $\mathbf{K'}$
states. We thus find that the noninteracting ground state of undoped
BLG possesses two pairs of {}``helical'' edge states with opposite
chiralities from different valleys. The ground state is characterized
by a sharp valley domain wall around $x=0$, i.e. at zero bias.

\section{\label{sec:Coherence-domain-walls}Interaction-induced valley kink:
Hartree-Fock treatment}

When the Coulomb interaction is incorporated, the system develops
a ferromagnetic nature.\cite{Fertig1994} A sharp domain wall between
two spin states or two valleys, as obtained in the non-interacting
ground-state described above, is not energetically favorable due to
its large cost in exchange energy. The competition between the single-particle
energy and the exchange energy gives rise to a lower energy state:
a smooth kink in the spin/valley degrees of freedom.

We focus on the situation shown in the center of Fig. \ref{fig:Quantum-Hall-energy},
where below the Fermi surface the filled energy states on the left
are dominated by the $\mathbf{K}$ valley, and on the right by the
$\mathbf{K'}$ valley. The Coulomb interaction modifies the sharp
domain wall to a smooth valley kink which can be described by a trial
wavefunction of the form\cite{Fertig2006} 
\begin{equation}
\left|\Psi\right\rangle =\prod_{X}\left(\cos\frac{\theta(X)}{2}C_{\mathbf{K}X}^{\dagger}+\sin\frac{\theta(X)}{2}e^{i\varphi}C_{\mathbf{K'}X}^{\dagger}\right)\left|0\right\rangle .\label{eq:trialwave}
\end{equation}
 Here $C_{\mathbf{K}X}^{\dagger}$ and $C_{\mathbf{K'}X}^{\dagger}$
create electrons in the levels $\varepsilon_{2}^{\mathbf{K}}(X)$
and $\varepsilon_{2}^{\mathbf{K'}}(X)$ respectively, which are closest
to the Fermi energy, and we assume that the $\varepsilon_{1}$ states
are pushed farther away from the Fermi level when the Coulomb interaction
is included. $\left|0\right\rangle $ denotes the vacuum state where
all lower states of negative single-particle energy are occupied,
and $\varphi$ is a constant parameter. The function $\theta(X)$
defines the valley profile of the domain wall varying from $0$ to
$\pi$: as shown on the far left of Fig. \ref{fig:Quantum-Hall-energy},
the filled state below the Fermi level is the $\mathbf{K}$ valley
state which corresponds to $\theta=0$; likewise on the far right
of the figure, the filled state below the Fermi level is the $\mathbf{K'}$
which corresponds to $\theta=\pi$. Eq. (\ref{eq:trialwave}) may
be regarded as a restricted Hartree-Fock approximation to the groundstate.
In the following, we use this to study the guiding center dependence
of $\theta(X)$, and determine the width of the domain wall.

The total Hamiltonian of the interacting electron system is 
\begin{equation}
H=H_{0}+H_{int},\label{eq:tH}
\end{equation}
 with single particle energy 
\begin{equation}
H_{0}={\displaystyle {\displaystyle \sum_{\tau X}\varepsilon_{\tau X}C_{\tau X}^{\dagger}C_{\tau X}}}\label{eq:single}
\end{equation}
 and interaction 
\begin{equation}
H_{int}=\frac{1}{2}\int\mathtt{d}\mathbf{r}\mathtt{d}\mathbf{r'}:\rho(\mathbf{r})V(\mathbf{r}-\mathbf{r'})\rho(\mathbf{r'}):.
\end{equation}
 Here $:(\cdot\cdot\cdot):$ indicates normal ordering. The density
operator is projected into the two states closest to the Fermi energy,
\begin{equation}
\rho(\mathbf{r})={\textstyle {\displaystyle \sum_{\tau,\tau',X,X'}{\displaystyle \phi_{\tau'X'}^{*}(\mathbf{r})\phi_{\tau X}(\mathbf{r})C_{\tau'X'}^{\dagger}C_{\tau X},}}}
\end{equation}
 with $\phi$ as defined in Eq. (\ref{eq:phi2}), $\tau^{(')}$ represents
the valley index $\mathbf{K}$ and $\mathbf{K'}$, and $V(\mathbf{r}-\mathbf{r'})=\frac{e^{2}}{\kappa l_{B}\left|\mathbf{r}-\mathbf{r'}\right|}$
is the Coulomb interaction among the electrons ($\kappa\sim5.2$ for
a suspended bilayer graphene).\cite{Gonzalez1999,Ghahari2011} We
apply the Hartree-Fock approximation to the total Hamiltonian in Eq.
(\ref{eq:tH}) and evaluate expectation values in the ground state
given by Eq. (\ref{eq:trialwave}).\textbf{ }The total Hamiltonian
can therefore be written as $H={\displaystyle {\textstyle \sum_{X}}H_{X}^{HF}}$,
where $H_{X}^{HF}$ is an effective $2\times2$ Hamiltonian for each
guiding center coordinate in the basis of $\left|\mathbf{K}X\right\rangle $
and $\left|\mathbf{K'}X\right\rangle $,

\begin{equation}
H_{X}^{HF}=\left[\begin{array}{cc}
\left(\varepsilon_{\mathbf{K}X}+J_{\mathbf{K}X,\mathbf{K}X}\right) & J_{\mathbf{K}X,\mathbf{K'}X}\\
J_{\mathbf{K}X,\mathbf{K'}X}^{*} & \left(\varepsilon_{\mathbf{K'}X}+J_{\mathbf{K'}X,\mathbf{K'}X}\right)
\end{array}\right].\label{eq:ME}
\end{equation}
 Here $\varepsilon_{\mathbf{K'}X}=-\varepsilon_{\mathbf{K}X}$ denote
the single particle energies given by Eq. (\ref{eq:e2}), and the
interaction terms are

\begin{eqnarray}
J_{\mathbf{K}X,\mathbf{K}X} & = & E_{H}-\frac{1}{2}\sum_{X'}\left\langle C_{\mathbf{K}X'}^{\dagger}C_{\mathbf{K}X'}\right\rangle V_{X,X'},\label{jkxkx}\\
J_{\mathbf{K'}X,\mathbf{K'}X} & = & E_{H}-\frac{1}{2}\sum_{X'}\left\langle C_{\mathbf{K'}X'}^{\dagger}C_{\mathbf{K'}X'}\right\rangle V_{X,X'},\\
J_{\mathbf{K}X,\mathbf{K'}X} & = & -\frac{e^{i\varphi}}{2}\sum_{X'}\left\langle C_{\mathbf{K'}X'}^{\dagger}C_{\mathbf{K}X'}\right\rangle V_{X,X'},\label{eq:Jgap}
\end{eqnarray}
 in which 
\[
\left\langle C_{\mathbf{K}X'}^{\dagger}C_{\mathbf{K}X'}\right\rangle =\cos^{2}\frac{\theta(X')}{2},
\]

\[
\left\langle C_{\mathbf{K'}X'}^{\dagger}C_{\mathbf{K'}X'}\right\rangle =\sin^{2}\frac{\theta(X')}{2},
\]

\[
\left\langle C_{\mathbf{K'}X'}^{\dagger}C_{\mathbf{K}X'}\right\rangle =\cos\frac{\theta(X')}{2}\sin\frac{\theta(X')}{2}.
\]
 $E_{H}=\sum_{X'}V_{X',X,X,X'}$, where the integral $V_{X_{1},X_{2},X_{3},X_{4}}$
is defined in Eqs. (\ref{eq:vtotal}) and (\ref{eq:v1-1}-\ref{eq:v4}),
denotes the Hartree contribution to the single particle energies.
Although $E_{H}$ is formally divergent, in practice it is canceled
by interactions with a uniform neutralizing background, which is not
explicitly included in our Hamiltonian. The exchange interaction matrix
element $V_{X,X'}\equiv V_{X',X,X',X}$ is given by

\begin{widetext} 
\begin{equation}
V_{X,X'}=\frac{V_{0}}{L_{y}}e^{-\frac{\left(X-X'\right)^{2}}{4}}\left\{ U_{0}(X-X')K_{0}\left[\frac{\left(X-X'\right)^{2}}{4}\right]+U_{1}(X-X')K_{1}\left[\frac{\left(X-X'\right)^{2}}{4}\right]\right\} ,\label{eq:coulombintegral}
\end{equation}

\end{widetext}where $V_{0}=\frac{e^{2}}{32\kappa l_{B}N_{h}^{4}}$,
$K_{n}$ are the modified Bessel functions which are localized at
$\left|X-X'\right|<1$ , and $U_{0,1}$ denote polynomial functions
of $\left(X-X'\right)$ as described in Appendix \ref{sec:Evaluation-of-the}.

As shown in Eq. (\ref{eq:ME}), the Coulomb interaction introduces
off-diagonal exchange terms which open a gap, and yields a smooth
domain wall as described by Eq. (\ref{eq:trialwave}). The trial wavefunction
$\left|\Psi\right\rangle $ obeys the eigenvalue equation 
\[
H_{X}^{HF}\left|\Psi\right\rangle =\varepsilon\left|\Psi\right\rangle .
\]
 Using Eqs. (\ref{eq:trialwave}) and (\ref{eq:ME}), this yields

\begin{widetext}

\begin{equation}
\left[\begin{array}{cc}
\left(\varepsilon_{\mathbf{K}X}+B_{X}\right) & \triangle_{X}\\
\triangle_{X}^{*} & -\left(\varepsilon_{\mathbf{K}X}+B_{X}\right)
\end{array}\right]\left[\begin{array}{c}
\cos\frac{\theta(X)}{2}\\
\sin\frac{\theta(X)}{2}
\end{array}\right]=\left(\varepsilon-A\right)\left[\begin{array}{c}
\cos\frac{\theta(X)}{2}\\
\sin\frac{\theta(X)}{2}
\end{array}\right],\label{eq:BCSgap}
\end{equation}

\end{widetext}where we define 
\begin{eqnarray*}
\triangle_{X} & = & J_{\mathbf{K}X,\mathbf{K'}X},\\
A & = & E_{H}+\frac{1}{2}\left(J_{\mathbf{K}X,\mathbf{K}X}+J_{\mathbf{K'}X,\mathbf{K'}X}\right),\\
B_{X} & = & \frac{1}{2}\left(J_{\mathbf{K}X,\mathbf{K}X}-J_{\mathbf{K'}X,\mathbf{K'}X}\right).
\end{eqnarray*}
 This yields the relation 
\begin{equation}
\left|\triangle_{X}\right|=\left(\varepsilon_{\mathbf{K}X}+B_{X}\right)\tan\theta(X),\label{eq:Trirelation}
\end{equation}
 and consequently 
\begin{equation}
\sin\theta(X)=\frac{\left|\triangle_{X}\right|}{\sqrt{\left|\triangle_{X}\right|^{2}+\left(\varepsilon_{\mathbf{K}X}+B_{X}\right)^{2}}}\label{eq:sin}
\end{equation}
 and 
\begin{equation}
\cos\theta(X)=\frac{\left(\varepsilon_{\mathbf{K}X}+B_{X}\right)}{\sqrt{\left|\triangle_{X}\right|^{2}+\left(\varepsilon_{\mathbf{K}X}+B_{X}\right)^{2}}}.\label{eq:cos}
\end{equation}
 The quantities $\left|\triangle_{X}\right|$ and $B_{X}$ must be
determined self-consistently.

Using Eq. (\ref{eq:sin}) for $\sin\theta(X)$ in Eq. (\ref{eq:Jgap})
and replacing the sums over $X'$ by integrals, we obtain two coupled
gap equations,

\begin{equation}
\left|\triangle_{X}\right|=-\frac{L_{y}}{4\pi}\int\mathtt{d}X'\frac{\left|\triangle_{X'}\right|}{\sqrt{\left|\triangle_{X'}\right|^{2}+\left(\varepsilon_{\mathbf{K}X'}+B_{X'}\right)^{2}}}V_{X,X'},
\end{equation}

\begin{equation}
B_{X}=-\frac{L_{y}}{4\pi}\int\mathtt{d}X'\frac{\varepsilon_{\mathbf{K}X'}+B_{X'}}{\sqrt{\left|\triangle_{X'}\right|^{2}+\left(\varepsilon_{\mathbf{K}X'}+B_{X'}\right)^{2}}}V_{X,X'}.
\end{equation}
 When $X$ is near the center (i.e. $X\ll w$), we can assume $\triangle_{X'}\sim\triangle_{X}$
and $B_{X}\sim B_{X'}$. Since $\triangle_{X}\left(\triangle_{X}\neq0\right)$
is maximum at $X=0$, and $B_{X}\rightarrow0$ at $X=0$, an approximate
solution to the gap equations takes the form 
\begin{equation}
\triangle_{X}\approx\sqrt{\triangle_{0}^{2}-\left(\eta X+B_{X}\right)^{2},}\label{eq:UltimateGap}
\end{equation}
 where 
\[
\triangle_{0}=V_{0}\sqrt{\frac{\pi}{8}}\left(8+8\beta^{2}+3\beta^{4}\right),
\]
 with $\beta=-\gamma_{1}/\omega_{c}$, and 
\[
\eta=\frac{\left(\gamma_{1}^{2}-\omega_{c}^{2}\right)V}{\left(\gamma_{1}^{2}+\omega_{c}^{2}\right)w},
\]
 which is the effective slope of the position dependent perpendicular
bias; 
\begin{eqnarray}
B_{X} & \sim & -\frac{\eta}{\left(1+f(\beta)\right)}X,\label{eq:partB}
\end{eqnarray}
 with $f(\beta)=\left(8+8\beta^{2}+3\beta^{4}\right)/\left(8+8\beta^{2}+6\beta^{4}\right)$.

\begin{figure}
\includegraphics[width=1\columnwidth]{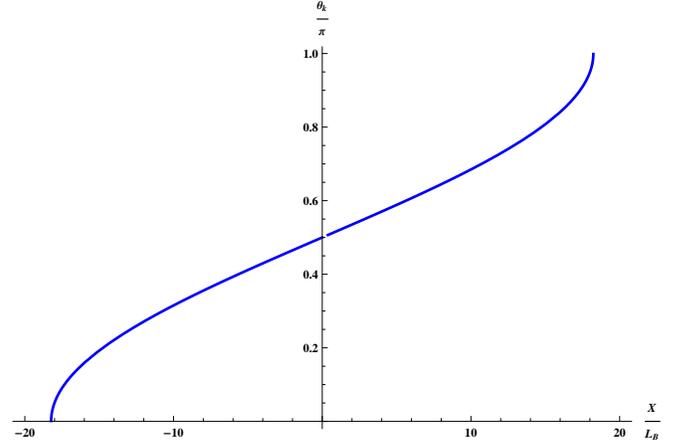}

\caption{\label{fig:Variational-angle-}The guiding center dependence of $\theta$.
The parameters are chosen as follows: $V=200$ meV, $l_{B}=10$ nm,
$w=20\: l_{B}$, $\gamma_{1}/\omega_{c}=3$.}
\end{figure}

Combining Eqs. (\ref{eq:UltimateGap}) and (\ref{eq:partB}), we obtain
\begin{equation}
\left|\triangle_{X}\right|\sim\sqrt{\triangle_{0}^{2}-\eta^{2}\left(1-\frac{1}{1+f(\beta)}\right)^{2}X^{2}}.
\end{equation}
 Substituting this into Eq. (\ref{eq:Trirelation}), this yields an
expression for $\theta(X)$. In Fig. \ref{fig:Variational-angle-}
we plot this for sample parameters as listed in the caption of the
figure. The width of the valley domain wall may be estimated to be
\[
d_{DW}\sim\frac{\triangle_{0}}{\eta}\sim\frac{\triangle_{0}}{V}w
\]
 which is in general dependent on the ratio of the maximal Coulomb
gap $\triangle_{0}$ (which has magnetic field dependence) to the
applied bias and the separation of the two opposing polarity gates.
$\theta(X)$ is almost linear in $X$ at the center and curves up
on the sides where the approximations of $\triangle_{X'}\sim\triangle_{X}$
and $B_{X}\sim B_{X'}$ are no longer valid. It is apparent that the
width of the kink can be tuned by the interplay between the magnetic
field and gate electric fields. We expect that an exact minimization
of the trial wavefunction would yield a very similar result near the
center of the DW, but the singularities in the slopes near the edges
would be smoothed out. 

\section{\label{sec:Interlayer-charge-density}Interlayer charge density pattern}

Here we propose a possible measurement to visualize the valley-kink
domain wall derived in the previous section. We start by projecting
the density operator of bilayer graphene into its four sublattices,
i.e. $\rho(\mathbf{r})={\displaystyle \sum_{\mu}\rho_{\mu}(\mathbf{r})},$
where 
\[
\rho_{\mu}(\mathbf{r})={\displaystyle \sum_{\tau,\tau'X,X'}\phi_{\mu\tau X}^{*}(\mathbf{r})\phi_{\mu\tau'X'}(\mathbf{r})C_{\tau X}^{\dagger}C_{\tau'X}}
\]
 in which $\mu$ represents the four sublattices $A$, $B$, $\tilde{A}$,
and $\tilde{B}$, and $\phi_{\mu}$ represents the $\mu$th component
of $\phi_{\mathbf{K}X}$ defined in Eq. (\ref{eq:phi2}).

\begin{figure}[t]
\begin{centering}
\includegraphics[width=1\columnwidth]{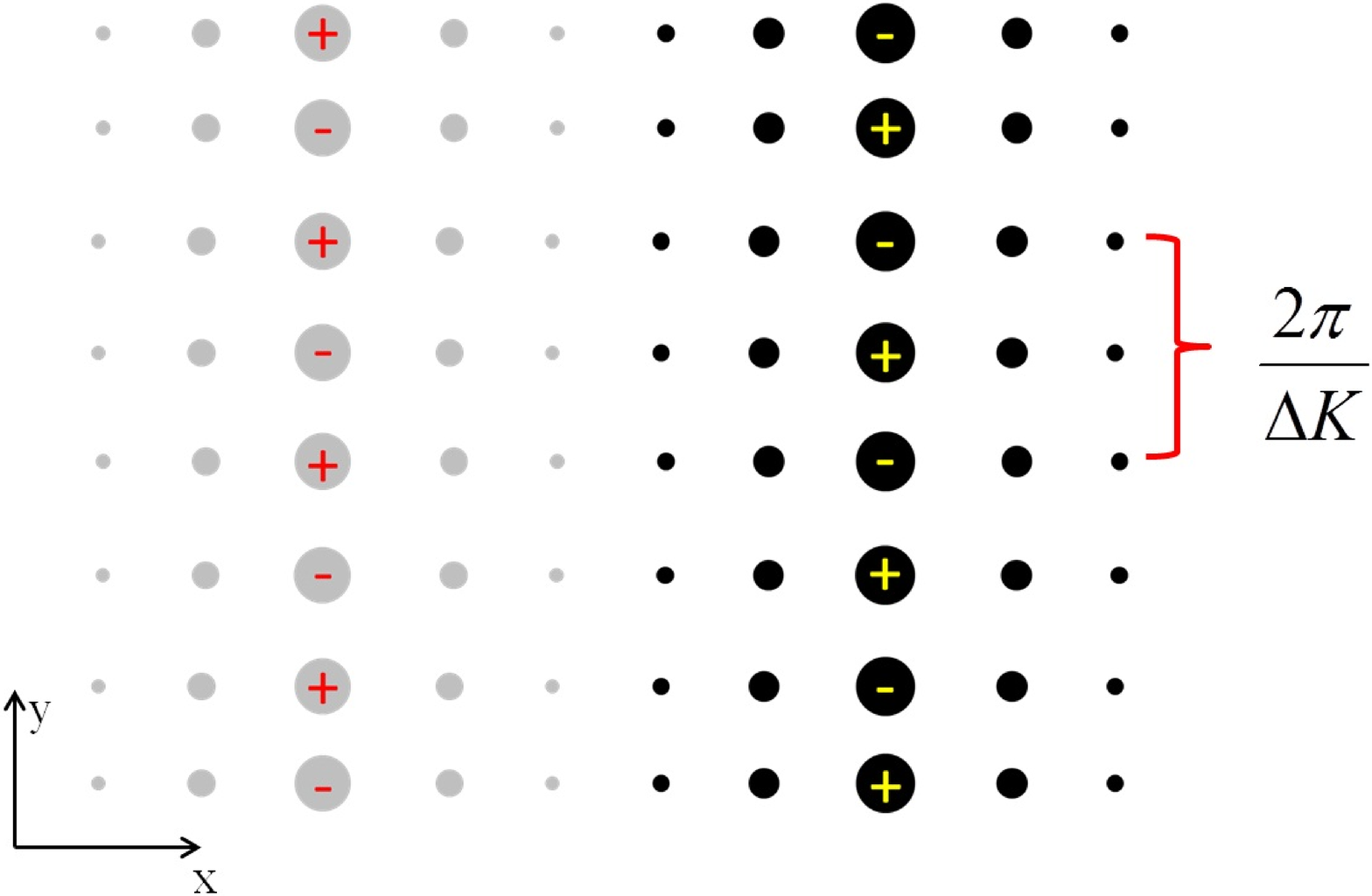} \vspace{8mm}
 \includegraphics[width=1\columnwidth]{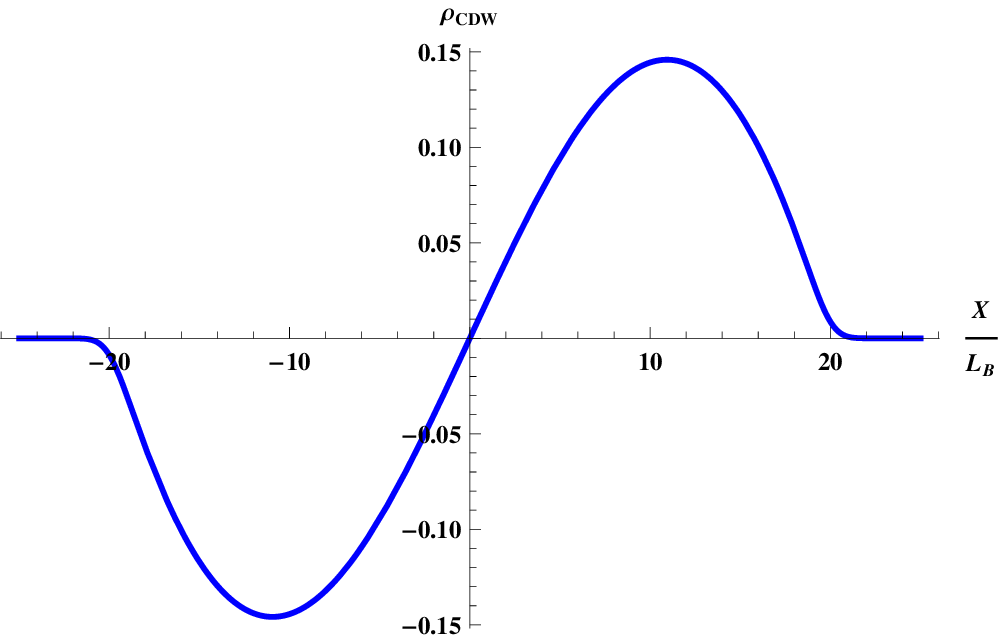} 
\par\end{centering}

\caption{\label{fig:Charge-density-wave}Interlayer charge density pattern
in a BLG domain wall. Upper panel: a rapid oscillation along the y
direction with wave vector $\triangle\mathbf{K}=\mathbf{K}-\mathbf{K'}$,
and a dipolar profile along the $x$ direction. Lower panel: at $y=0$
, the interlayer charge density pattern $\rho_{CDW}(x,0)$ the thick
blue curve is obtained from numerical integral. The parameters chosen
here are: $V=200$ meV, $l_{B}=10$ nm, $w=20\: l_{B}$, $\gamma_{1}/\omega_{c}=3$,
$\varphi=0$.}
\end{figure}

Using Eq. (\ref{eq:trialwave}), the expectation value of the density
on sublattice $\mu$ is

\begin{eqnarray}
\left\langle \rho_{\mu}(\mathbf{r})\right\rangle  & = & \left\langle \Psi\right|\rho_{\mu}(\mathbf{r})\left|\Psi\right\rangle \label{eq:subroh}\\
 & = & {\displaystyle \sum_{X}\cos^{2}\frac{\theta(X)}{2}R_{\mathbf{KK}X}^{\mu}(\mathbf{r})+\sin^{2}\frac{\theta(X)}{2}R_{\mathbf{K'K'}X}^{\mu}(\mathbf{r})}\nonumber \\
 &  & +e^{i\varphi}\cos\frac{\theta(X)}{2}\sin\frac{\theta(X)}{2}\left[R_{\mathbf{KK'}X}^{\mu}(\mathbf{r})+R_{\mathbf{K'K}X}^{\mu}(\mathbf{r})\right],\nonumber 
\end{eqnarray}
 where 
\[
R_{\tau\tau'X}^{\mu}(\mathbf{r})=\phi_{\mu\tau X}^{*}(\mathbf{r})\phi_{\mu\tau'X}(\mathbf{r}).
\]
 The last term of Eq. (\ref{eq:subroh}) indicates interference between
the $\mathbf{K}$ and $\mathbf{K'}$ valleys. In the valley transition
region, an interference pattern would therefore be manifested by the
charge density difference between the top ($t$) and the bottom ($b$)
layer of the BLG,

\begin{eqnarray}
\triangle\rho(x,y) & = & \rho_{t}-\rho_{b},\nonumber \\
 & = & \left(\rho_{A}(\mathbf{r})+\rho_{B}(\mathbf{r})\right)-\left(\rho_{\tilde{A}}(\mathbf{r})+\rho_{\tilde{B}}(\mathbf{r})\right),\\
 & = & \rho_{0}(x)+\rho_{CDW}(x)\cos\left(\Delta\mathbf{K}y+\varphi\right).\label{eq:CDW-1}
\end{eqnarray}
 Here 
\begin{eqnarray}
\rho_{0}(x) & = & \frac{1}{N_{h}^{2}}\sum_{X}\cos\theta(X)\times\\
 &  & \left\{ \beta^{2}\Phi_{1}^{2}(x-X)-\left[1-\alpha(X)^{2}\right]\Phi_{0}^{2}(x-X)\right\} ,\nonumber 
\end{eqnarray}

\begin{equation}
\rho_{CDW}(x)=\frac{4\gamma_{1}V}{N_{h}^{2}\left(\gamma_{1}^{2}+\omega_{c}^{2}\right)w}\sum_{X}X\Phi_{0}^{2}(x-X)\sin\theta(X),\label{eq:cdp}
\end{equation}
 with $\triangle\mathbf{K}=\mathbf{K}-\mathbf{K'}$, 
\[
\alpha(X)=\frac{2\gamma_{1}VX}{\left(\gamma_{1}^{2}+\omega_{c}^{2}\right)w},
\]
 and the normalization factor $N_{h}\sim\sqrt{1+\beta^{2}}$. The
first term in Eq. (\ref{eq:CDW-1}) represents the average charge
density difference between the top and the bottom layers, while the
second term describes a charge density wave through which we see a
rapid oscillation along the $y$ direction with wave vector $\triangle\mathbf{K}$.
The resulting interlayer charge density pattern is shown in Fig. \ref{fig:Charge-density-wave},
wherein the upper panel displays an intervalley interference along
the $y$ direction and a dipolar charge profile along the $x$ direction.
Across $x=0$, the interlayer charge density pattern shows an interesting
antisymmetric amplitude which is due to the switch in polarity of
the potential profile. 

\section{\label{sec:Concluding-remarks}Concluding remarks}

We have proposed an experimental setup to realize a collective, smooth
kink in the valley degrees of freedom for bilayer graphene at $\nu=0$
in the presence of a spatially antisymmetric bias field. The width
of the kink is determined by an interplay between the magnetic field
and the gate electric fields. We predict a potentially measurable
interlayer charge density pattern to visualize this resulting electronic
structure. According to Eq. (\ref{eq:cdp}), the amplitude of the
charge density pattern can be tuned by the ratio of $V$ to $\gamma_{1}$.
This pattern is possibly accessible to measurement, e.g. by an STM
probe.

The above results assume that the Zeeman splitting of the real spin
is negligibly small compared to the maximal valley splitting set by
the gate voltage. We note that for sufficiently strong magnetic fields
where the real spin is resolved, two distinct crossing points appear
in the non-interacting spectrum at zero energy, separated by a finite
distance in real space. Consequently, a more complex double-kink pattern
is expected to form in the interacting groundstate, which can be viewed
as a pair of DW's with a mutual interaction which is tunable by the
gate-voltage. This case will be studied elsewhere.\cite{Huang}

We conclude with speculations about the collective electronic transport
behavior of this system. In analogy with what happens with a spin
DW at the edge of single layer graphene at $\nu=0$,\cite{Fertig2006,Shimshoni2009}
we expect the DW to carry valley currents which can lead to a valley
QHE. Unlike the single layer case, the non-interacting energy spectrum
for the bilayer structure we consider has two pairs of states crossing
the Fermi level, although one has much greater slope than the other.
For long length scales, and for the purposes of static properties,
in a first approximation one may ignore the higher energy states as
we have done in this study. However, very close to $x=0$ the second
pair of internal edge states will likely give the charge density profile
further structure as they approach zero energy. More importantly,
these extra states crossing the Fermi energy open a second current-carrying
channel, which will affect the transport properties of the system.
An interesting set of questions in this regard is how the second channel
couples to the first, in particular if they can be regarded as independent
channels or if they are locked together by Coulomb interactions. Finally,
we note that in the case where the splitting of real-spin is appreciable,
the two coupled DWs are likely to support a quasi 1D collective mode
characterized by a ladder-like dynamics. We leave these questions
for future research. 
\begin{acknowledgments}
We acknowledge useful discussion with E. Andrei, V. Mazo and A. Yacoby.
We thank financial support from the US-Israel Binational Science Foundation
(BSF) through Grant No. 2008256 , and by the US National Science Foundation
(NSF) through Grant No. DMR1005035. H. A. F. and E. S. are grateful to the hospitality of the Aspen Center for Physics (NSF 1066293), where part of this work was carried out.
\end{acknowledgments}
\appendix
%dummy comment inserted by tex2lyx to ensure that this paragraph is not empty
%dummy comment inserted by tex2lyx to ensure that this paragraph is not empty
%dummy comment inserted by tex2lyx to ensure that this paragraph is not empty
%dummy comment inserted by tex2lyx to ensure that this paragraph is not empty
%dummy comment inserted by tex2lyx to ensure that this paragraph is not empty

\section{\label{sec:Evaluation-of-the}Evaluation of the coulomb integrals}

Using Eqs. (\ref{eq:phi1}) and (\ref{eq:phi2}) we write the Coulomb
integral as follows: 
\begin{equation}
V_{X_{1},X_{2},X_{3},X_{4}}=C_{1}V^{(1)}+C_{2}V^{(2)}+C_{3}\left(V^{(3)}+V^{(4)}\right),\label{eq:vtotal}
\end{equation}
 where up to corrections of order $(V/\gamma_{1})^{2}$ 
\begin{eqnarray*}
C_{1} & \approx & \frac{e^{2}}{L_{y}^{2}\kappa l_{B}N_{h}^{4}},\\
C_{2} & \approx & \frac{e^{2}\beta^{4}}{L_{y}^{2}\kappa l_{B}N_{h}^{4}},\\
C_{3} & \approx & \frac{e^{2}\beta^{2}}{L_{y}^{2}\kappa l_{B}N_{h}^{4}},
\end{eqnarray*}
 with $\beta=-\gamma_{1}/\omega_{c}$, and

\begin{widetext} 
\begin{eqnarray}
V^{(1)} & = & \int\mathtt{d}\mathbf{r}\mathtt{d}\mathbf{r'}e^{i\left(X_{2}-X_{1}\right)\left(y-y'\right)}\Phi_{0}^{*}(x-X_{1})\Phi_{0}^{*}(x'-X_{2})V(\mathbf{r}-\mathbf{r'})\Phi_{0}(x'-X_{3})\Phi_{0}(x-X_{4}),\label{eq:v1-1}\\
V^{(2)} & = & \int\mathtt{d}\mathbf{r}\mathtt{d}\mathbf{r'}e^{i\left(X_{2}-X_{1}\right)\left(y-y'\right)}\Phi_{1}^{*}(x-X_{1})\Phi_{1}^{*}(x'-X_{2})V(\mathbf{r}-\mathbf{r'})\Phi_{1}(x'-X_{3})\Phi_{1}(x-X_{4}),\label{eq:v2}\\
V^{(3)} & = & \int\mathtt{d}\mathbf{r}\mathtt{d}\mathbf{r'}e^{i\left(X_{2}-X_{1}\right)\left(y-y'\right)}\Phi_{0}^{*}(x-X_{1})\Phi_{1}^{*}(x'-X_{2})V(\mathbf{r}-\mathbf{r'})\Phi_{1}(x'-X_{3})\Phi_{0}(x-X_{4}),\label{eq:v3}\\
V^{(4)} & = & \int\mathtt{d}\mathbf{r}\mathtt{d}\mathbf{r'}e^{i\left(X_{2}-X_{1}\right)\left(y-y'\right)}\Phi_{1}^{*}(x-X_{1})\Phi_{0}^{*}(x'-X_{2})V(\mathbf{r}-\mathbf{r'})\Phi_{0}(x'-X_{3})\Phi_{1}(x-X_{4}).\label{eq:v4}
\end{eqnarray}

\end{widetext} In particular, for $X_{1}=X_{3}=X'$ and $X_{2}=X_{4}=X$
this yields the exchange interaction terms $V_{X,X'}$. {}As an example,
Eq. (\ref{eq:v1-1}) may be written explicitly in the form

\begin{widetext}

\begin{eqnarray*}
V^{(1)} & = & \frac{1}{\pi}\int\mathtt{d^{2}}r\mathtt{d^{2}}r'e^{i\left(X-X'\right)\left(y-y'\right)}\frac{\exp\left\{ -\frac{1}{2}\left[\left(x-X\right)^{2}+\left(x'-X'\right)^{2}+\left(x'-X\right)^{2}+\left(x-X'\right)^{2}\right]\right\} }{\sqrt{\left(x-x'\right)^{2}+\left(y-y'\right)^{2}}}.
\end{eqnarray*}

\end{widetext}

To evaluate this, we change variables to difference and center coordinates
$\tilde{x}=\left(x-x'\right)$, $\tilde{y}=y-y'$, $x_{c}=\frac{\left(x+x'\right)}{2}$,
and $y_{c}=\frac{\left(y+y'\right)}{2}$, and first integrate over
$x_{c}$ and $y_{c}$, to obtain

\begin{equation}
V^{(1)}=\frac{L_{y}\sqrt{2\pi}}{4\pi}e^{-\frac{\left(X-X'\right)^{2}}{2}}\int\mathtt{d}\tilde{x}\mathtt{d}\tilde{y}\frac{e^{i\left(X-X'\right)\tilde{y}}e^{-\frac{\tilde{x}^{2}}{2}}}{\sqrt{\tilde{x}^{2}+\tilde{y}^{2}}}.\label{eq:v1}
\end{equation}

We then use the 2D Fourier transform of the Coulomb potential 
\begin{eqnarray}
\frac{1}{\sqrt{\tilde{x}^{2}+\tilde{y}^{2}}} & = & \frac{1}{2\pi}\int\mathtt{d}^{2}k_{2}\frac{e^{-i\mathbf{k_{2}}\cdot\triangle\mathbf{r}}}{k_{2}},\label{eq:coulomb}
\end{eqnarray}
 to obtain\cite{Jeffrey2007}

\begin{eqnarray}
V^{(1)} & = & \frac{L_{y}}{2}e^{-\frac{\left(X-X'\right)^{2}}{2}}\int\mathtt{d}k_{2x}\frac{e^{-\frac{k_{2x}^{2}}{2}}}{\sqrt{k_{2x}^{2}+\left(X-X'\right)^{2}}},\nonumber \\
 & = & \frac{L_{y}}{2}e^{-\frac{\left(X-X'\right)^{2}}{4}}K_{0}\left[\frac{\left(X-X'\right)^{2}}{4}\right].
\end{eqnarray}

Similarly we can evaluate $V^{(2)}$, $V^{(3)}$, and $V^{(4)}$ and
obtain the final expression for the Coulomb integral in Eq. (\ref{eq:coulombintegral}),
with

\begin{widetext}

\begin{eqnarray}
U_{0}(X-X') & = & \beta^{4}\left[\left(X-X'\right)^{4}-4\left(X-X'\right)^{2}+8\right]-4\beta^{2}\left[\left(X-X'\right)^{2}-4\right]+8,\\
U_{1}(X-X') & = & \beta^{2}\left(X-X'\right)^{2}\left\{ \beta^{2}\left[\left(X-X'\right)^{2}-2\right]-4\right\} .
\end{eqnarray}

\end{widetext}

\end{document}